\documentclass[aps,pra,nofootinbib,longbibliography,notitlepage,twocolumn]{revtex4}
\usepackage{etex}
\usepackage{mathrsfs}
\usepackage{amsmath,amssymb,amsthm}
\newcommand{\ba}{\[\begin{aligned}}
\newcommand{\ea}{\end{aligned}\]}
\usepackage[colorlinks=true,citecolor=blue,urlcolor=blue]{hyperref}
\usepackage[pdftex]{graphicx}
\usepackage{times,txfonts}
\usepackage{braket}
\usepackage{color}
\usepackage{xcolor}
\usepackage{natbib}
\usepackage{amsmath,blkarray}
\usepackage{mathtools}
\usepackage{latexsym}
\usepackage{tabularx, booktabs}
\usepackage{graphics,epstopdf}
\usepackage{graphicx}
\usepackage{float}
\usepackage{multirow}
\usepackage{graphicx}
\usepackage{amsfonts}
\usepackage{cleveref}
\usepackage{comment}
\usepackage{subcaption}
\usepackage{pifont}
\usepackage{caption}
\usepackage{booktabs}  
\usepackage{pifont}   
\usepackage[table]{xcolor} 
\newcommand{\RNum}[1]{\uppercase\expandafter{\romannumeral #1\relax}}
\usepackage{color,soul}

\newcommand{\be}{\begin{equation}}
\newcommand{\ee}{\end{equation}}

\begin{document}

\title{Probing the extreme quantum violation of the  Leggett-Garg inequality}

\author{Akarshit Baranwal}
\email{akarshitbaranwal@gmail.com}
\affiliation{Department of Physics, Indian Institute of Technology Hyderabad, Kandi, Sangareddy, Telengana-502284, India.}

\author{Vishnu Purushothaman}
\email{vishnu_purushothaman@icloud.com }
\affiliation{Department of Physics, Indian Institute of Technology Hyderabad, Kandi, Sangareddy, Telengana-502284, India.}

\author{A. K. Pan\textsuperscript{*}}
	\email{akp@phy.iith.ac.in(Corresponding author)}
	\affiliation{Department of Physics, Indian Institute of Technology Hyderabad, Kandi, Sangareddy, Telengana-502284, India.}

\begin{abstract}

The Leggett-Garg (LG) inequalities were formulated for testing the incompatibility between the classical worldview of macrorealism and quantum theory.  By analogy with the Tsirelson bound for Bell inequalities, the maximal quantum violation of LG inequalities is commonly referred to as the temporal Tsirelson bound (TTB). Nevertheless, several studies have claimed that, unlike the Tsirelson bound, the TTB can be exceeded within quantum theory.  In particular, by employing degeneracy-breaking measurement in [\href{ https://doi.org/10.1103/PhysRevLett.113.050401}{Phys. Rev. Lett. \textbf{113}, 050401 (2014)}] and   by using the superposition of unitaries in [\href{https://doi.org/10.1103/vydp-9qqq}{Phys. Rev. Lett. \textbf{135}, 220202 (2025)}], it is argued that the TTB can be violated. In this work, we critically examine the reasoning underlying these purported TTB violations and show that the corresponding arguments suffer from conceptual shortcomings. We further show that once these conceptual issues are addressed appropriately, the macrorealist bound of LG inequality reaches the algebraic maximum, and hence no violation of the TTB is obtained. In particular, we demonstrate that the proposed TTB violations rely on frameworks that go beyond the standard LG framework, and therefore are of no relevance for testing the incompatibility between macrorealism and quantum theory.

	\end{abstract}
	\maketitle
    \section{Introduction}
The notion of macrorealism posits that a macroscopic object in our everyday world possesses
a definite state, even though it is not precisely
known. However, it remains a debatable question as to how such a macrorealistic view is compatible with the formalism of quantum theory.  This question was first posed by  Schr\"odinger \cite{sch} through his famous cat experiment. Since then, a number of approaches have been made \cite{zeh, zur,ghc86, ghc03,bruk} to shed light on the compatibility between macrorealism and quantum theory. A practical approach is to demonstrate the quantum coherence of large objects in an interference experiment \cite{arndt,gerlich,Schrinski,ped}. However, the aforementioned attempts do not directly address the fundamental question of how a realist view of the macroscopic classical world emerges from the framework of quantum theory.

Motivated by Bell's idea of local realism \cite{bell64, triel}, Leggett and Garg  \cite{leggett85} first provided a refined notion of macrorealism, which consists of two main assumptions \cite{leggett85, leggett02, leggett08}.  (i) \emph{Macrorealism }{\emph per se} (MRps): If a macroscopic system has two or more states available to it, it remains in one of them at any instant of time. (ii) \emph{Non-invasive measurability (NIM)}: It is possible, at least in principle, to determine the state of the system without affecting the system itself and its subsequent dynamics. They proposed \cite{leggett85} a set of inequalities (known as LG inequalities) that are valid in a macrorealistic model, satisfying the above assumptions. It is widely studied and demonstrated \cite{Emary13, halli16, halli17, halli19,halli22, wild, maroney, clemente15, clemente16, akp23, weak value and MR akp20, budroni23, budroni13, budroni14, swati17, pan17, kumari18, pan18, halli19a,brukner14, halli20} that such inequalities are violated by quantum theory, establishing the incompatibility between macrorealism and quantum theory.  Experimental tests have also been performed \cite{Goggin, robens, knee, usha25, laflamme, tusun20, tusun22, wang17, zhan23, pala,sponar, Zhang} to demonstrate the violation of LG inequalities. However, there remains a loophole \cite{wild, maroney, clemente15, clemente16, halli17, halli19, pan17, akp23, budroni23, weak value and MR akp20, halli19a,brukner14, halli22, halli20} arising from the question of salvaging MRps by abandoning the NIM assumption. It remains a debatable issue, and no clear consensus on this point has been achieved yet.

The present work concerns the enhanced quantum violation of macrorealism. By drawing the analogy of the Tsirelson bound \cite{triel} of the Bell inequality \cite{bell64}, the maximum quantum violation of the LG inequalities is termed TTB.  Budroni and Emary \cite{budroni14} have argued that TTB can be violated if a degeneracy-breaking measurement is used for state reduction. Very recently, considering the evolution between two measurements governed by the superposition of unitaries, Chatterjee \emph{et al.,} \cite{usha25} claimed to show the violation of TTB. In this paper, we show that there are conceptual gaps in those arguments \cite{budroni14, usha25} as they differ from the conventional LG framework. We show that, due to the degeneracy-breaking measurement employed in \cite{budroni14}, the derivation of three correlations in the LG functional is effectively performed for two different states. We further show that the three correlations in the LG functional in \cite{usha25} are effectively evaluated for four observables instead of three observables in the conventional LG framework. We demonstrate that the theoretical frameworks introduced in \cite{budroni14, usha25} go beyond the standard LG framework.  Furthermore, we show that their macrorealist bounds differ from those of the standard LG framework  and that, following an appropriate revision of these macrorealist bounds, no violation of TTB can be observed.

\section{The LG framework}
\begin{figure}[htp]
    \centering
    \captionsetup{justification=centering}
    \includegraphics[
  width=1\columnwidth,
  trim=30 90 30 50,
  clip]
{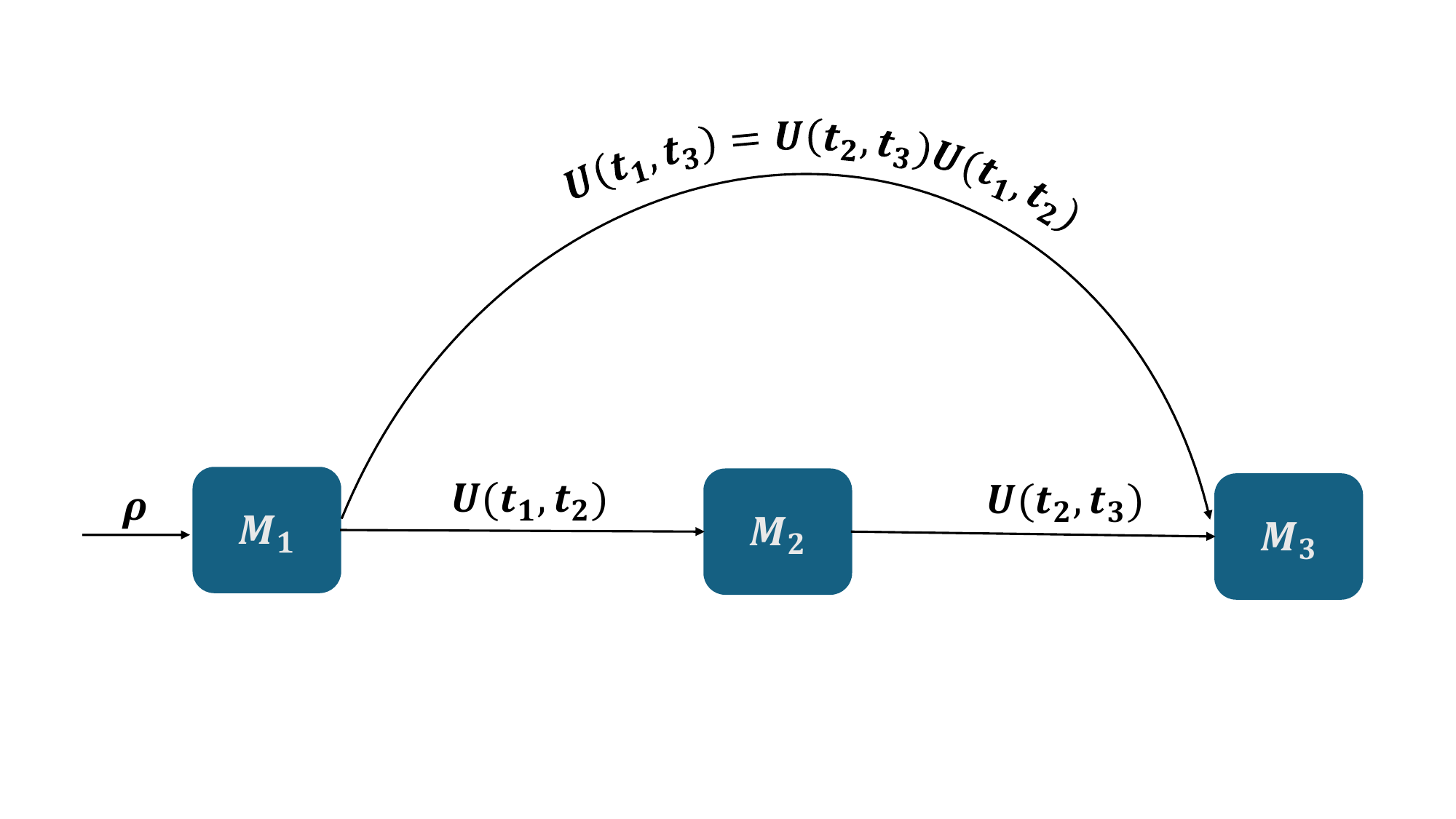}
    \caption{The conventional LG framework: Given  a state $\rho$, the  sequential measurements of observable $M_1$ is performed  at times  $t_1$, $ t_2$ and $ t_3$ with $t_3 > t_2 > t_1$. Here, $M_2$ and $M_3$ are unitarily evolved observables of $M_1$ at $t_2$ and $t_3$ respectively.}
    \label{fig:LG}
\end{figure}

In order to show this, let us first recapitulate the LG framework (as depicted in Fig. \ref{fig:LG}), which is quite well-defined. Any deviation from it may produce results that have bizarre consequences. In the three-time LG framework, a dichotomic observable $M_{1}$ is measured at three different times $t_1$, $t_2$ and $t_3$ with $t_3 >t_2 > t_1 $. This implies that the system unitarily evolves between two measurements. In the Heisenberg picture, this in turn implies the measurement of the observables  $M_{1}, M_{2}$ and $M_{3}$ corresponding to $t_1$, $t_2$ and $t_3$ respectively. The LG framework involves pair-wise sequential correlations at two different times. In the three-time LG scenario, for a given state $\rho$, the LG functional in quantum theory is defined as 


\begin{eqnarray}
    \label{lgi}
    K_{Q} = \langle {M_{1}} {M_{2}}\rangle _{\rho} +\langle {M_{2}} {M_{3}}\rangle _{\rho}-\langle {M_{1}} {M_{3}}\rangle _{\rho}
\end{eqnarray}

where  $\langle {M_{i}} {M_{j}}\rangle _{\rho} = \sum\limits_{m_{i},m_{j}=\pm1} m_{i} m_{j} p(m_{i}, m_{j})$ is the sequential correlation with $m_i m_j=\pm 1$ with $i,j=1,2,3$ and $i<j$. Here, $p(m_{i}, m_{j})$ is the sequential joint probability. It is shown \cite{leggett85, Emary13, swati17,pan18} that $(K_Q)_{max} = 1.5$.

It is important to note that, unlike the Bell inequality, in the LG scenario, the quantum LG functional in Eq.(\ref{lgi}) is first derived by considering the time evolution of $M_1$ to $M_2$ and then $M_3$. Further, the macrorealist bound of the LG functional $K$ is derived. Now, in a macrorealist model, corresponding to the density matrix $\rho$, one assumes the distribution $\mu(\lambda)$ where  $\lambda$ is the ontic state. Imposing the assumption of MRps and NIM on the ontic state, one derives the LG inequality $K\leq 1$. Therefore, a macrorealist model cannot reproduce quantum statistics, and thus establishes the incompatibility of macrorealism with quantum theory.

\subsection{Derivation of TTB}

We provide a very simple derivation of the optimal quantum value (the TTB) of the LG expression in Eq. (\ref{lgi}).  We note that for the sequential quantum measurement of two dichotomic observables $M_{i}$ and $M_{j}$ in Eq. (\ref{lgi}), one has \cite{pan18, akp23} 
\begin{eqnarray}
\label{seqq}
 \langle M_i, M_j \rangle_{\rho}  =\frac{ \operatorname{Tr} [\rho \{M_i, M_j\}]}{2}  
\end{eqnarray}
where $\{ \ , \}$ indicates the anticommutator. This relation holds for the coarse-grained L\"uders measurement (in this work we also term it as degeneracy-preserving measurement) of two dichotomic observables with eigenvalue $\pm 1$.

To derive the TTB of the LG functional $K_{Q}$ we consider a quantity $\Delta$ which is written as
\begin{eqnarray}
\label{eq2}
\Delta=\left(M_1-M_2 +M_3\right)^{2}.
\end{eqnarray}
By construction $\Delta\geq 0$. From Eq. (\ref{eq2}), we can then write

\begin{eqnarray}
\operatorname{Tr} [\rho \{M_1, M_2\}] + \operatorname{Tr} [\rho \{M_2, M_3\}]- \operatorname{Tr} [\rho \{M_1, M_3\}] \leq 3
\end{eqnarray}
Using Eq. (\ref{seqq}), the left-hand side can be identified as $2K_Q$ and hence  $K_{Q}\leq 1.5$. Therefore, $(K_{Q})_{max}=1.5$ is achieved when $\Delta =0$, which is the TTB of LG functional. This gives the condition on observables as $M_2 = M_1 + M_3$. 

Note that the derivation is irrespective of assuming dimension and is also devoid of assuming the state. This means that  $(K_{Q})_{max}=1.5$ can be obtained for any state in any arbitrary dimension if the relation $M_2 = M_1 + M_3$ is satisfied.   A qubit example is the following. $M_1 = \frac{\sigma_{z}}{2}+\frac{\sqrt{3}}{2}\sigma_{x}$, $M_2 = \sigma_{z}$,  and $M_3 = \frac{\sigma_{z}}{2}-\frac{\sqrt{3}}{2}\sigma_{x}$.

\subsection{Violation of TTB and motivation of present work}

The value of $K_{Q}$ exceeding TTB is then a signature of a higher degree of non-classicality. In comparison to the Bell scenario, a violation of the Tsirelson bound of the Bell inequality can only be achieved in post-quantum theories. However,  Budroni and Emary \cite{budroni14} have demonstrated that $K_{Q}$ can reach its algebraic maximum 3, violating the TTB. This is done by considering a degeneracy-breaking projective measurement, and the algebraic maximum can be achieved for a system with large dimension. Experimental demonstrations of such violation of TTB have also been reported \cite{laflamme, wang17, zhan23, tusun22}. Recently, Chatterjee \emph{et al.,} \cite{usha25} have also demonstrated violation of TTB. They considered that the evolution between two measurements is governed by the superposition of unitaries and argue that such evolution is the key to obtain the algebraic maximum $K_{Q}=3$. 

In this work, for better understanding, we first provide simpler examples to reproduce the results of \cite{budroni14, usha25}. In \cite{budroni14}, the observables were taken to be non-commuting. We show that for three observables which are mutually commuting in the coarse-grained  L\"uders measurement scenario, the TTB is violated if the fine-grained degeneracy-breaking measurement scheme is employed. We then show that the complex calculation using the linear superposition of unitaries used in \cite{usha25} is not necessary to obtain the violation of TTB.  A careful analysis shows that the scenarios invoked by implementing a degeneracy-breaking measurement \cite{budroni14} and by using the superposition of unitaries \cite{usha25}, they differ from the conventional LG framework. 
In such scenarios, the LG functionals differ from the one in Eq. (\ref{lgi}), and macrorealist bounds attain the algebraic maximum of these LG functionals.

 \section{ON the violation of TTB through degeneracy-breaking measurements}
 To make our paper self-contained, we give a brief description of degeneracy-breaking measurement \cite{heg,kathakali} and introduce elegant tools to derive sequential correlation in this scenario. To make our argument more interesting, instead of taking non-commuting observables in the LG scenario, we consider three mutually commuting observables. In a standard sequential measurement scenario involving coarse-grained degeneracy-preserving measurement, no quantum violation of LG inequality can occur for such mutually commuting observables. In contrast, by using fine-grained degeneracy-breaking measurement, we demonstrate that for the sequential measurement of aforementioned observables, not only the violation of the LG inequality but also the violation of the TTB occurs.

 \subsection{Fine-grained degeneracy-breaking measurement}

Let us briefly introduce the fine-grained degeneracy-breaking measurement. We restrict ourselves to the case of dichotomic observables. Consider an observable $A$ in an arbitrary $d$ dimension having eigenvalues $+1$ and $-1$ with degrees of degeneracies $n_+$ and $n_{-}$, respectively, where $n_{+}$ and $n_{-}$ can be arbitrary but $n_{+}+n_{-}=d$. Let $A_{+}^{k}=|\phi_{+}^{k}\rangle\langle\phi_{+}^{k}|$ and $A_{-}^{l}=|\phi_{-}^{l}\rangle\langle\phi_{-}^{l}|$ are the rank one projector corresponding to the eigen value $+1$ and $-1$ respectively, where $k\in\{1, 2, ..., n_{+}\}$ and $l\in\{1, 2, ..., n_{-}\}$. In addition, $A_{+}=\sum_{k=1}^{n_{+}}  A_{+}^{k}$ and  $A_{-}=\sum_{l=1}^{n_{-}}A_{-}^{l}$ are the rank $n_{+}$ and $n_{-}$ projectors, respectively, that satisfy $A=A_{+}-A_{-}$ and $A_{+}+A_{-}=\mathbb{I}_{d}$.

Note that the degeneracy-preserving and degeneracy-breaking projective measurements are originally known as the L\"uders and von Neumann measurements respectively. Now, for coarse-grained degeneracy-preserving projective measurement, the reduced density matrix can be written as

\begin{align}\label{rho_dp}
    \rho_{dp}={A_{+}\rho A_{+}}+{A_{-}\rho A_{-}}
\end{align}
where $\rho$ is the initial density matrix of the system. For fully fine-grained degeneracy-breaking projective measurements, the reduced density matrix can be written as
\begin{align}\label{rho_db}
    \rho_{db}=\sum_{k} A_{+}^{k}\rho A_{+}^{k}+\sum_{l}A_{-}^{l}\rho A_{-}^{l}
    \end{align}
In general, $\rho_{db}$ is not equivalent to $\rho_{dp}$ unless $d=2$ when there is no degeneracy. 

Note that in degeneracy-preserving projective measurement, only two projectors are used for state reduction, and in fully degeneracy-breaking measurement, $d$ number of projectors are used. However, any number of projectors  between $2$ and $d$ can be used for state reduction, which can be termed partial degeneracy-breaking measurement. Note that for the derivation of  the statistics for a single observable, say $\langle  A\rangle$, the way the state reduction occurs does not have any effect.  However, it becomes effective when the sequential measurement is performed in the reduced state after the first measurement. 

Consider two dichotomic observables $ A$ and $ B$ in a $d$ dimensional system. The sequential correlation of $A$ and $B$ can be written as
\begin{equation}
\langle A B \rangle = \sum_{a,b = \pm} a b \ p(a b)
\end{equation}
The probability $p(a b)$ can be calculated considering the state reduction for the observable $ A$ for the number of projectors $g$ where \textcolor{blue}{$g\in \{2,\ldots, d\}$}.  For different values of $g$ the probability  $p(a b)$ will be different. Consequently, the value of $\langle  A B \rangle$ will depend on the value of $g$.  

To sharpen the contrast between degeneracy-breaking and preserving measurements, and for our present purpose, we consider a concrete example. Let the measurement $A$ be performed by using $g=4$, i.e., four projectors are used for state reduction, so that we write
\begin{align}
A  = A^{1}_{+} + A^{2}_{+} - A^{1}_{-} - A^{2}_{-}
\end{align}
with ${A}_{+}=A^{1}_{+} + A^{2}_{+}$ and ${A}_{-}=A^{1}_{-} + A_{-}^{2}$. We make a connection between the expectation values $\langle AB \rangle_{dp}$ and $\langle A B \rangle_{db}$, to demonstrate the difference.

For coarse-grained degeneracy-preserving projective  measurement, the joint probability can be written as
\begin{equation}
p_{dp}(++) = \operatorname{Tr}
\!\left[
\left(A_{+}
\rho
A_{+}\right)
B_{+}
\right].
\end{equation}
Similarly, the remaining three joint probabilities can also be obtained. Noting that $B_{\pm}=\left(\mathbb{I}_{d}\pm B\right)/2$, we get the following.
\begin{align}\label{Luder}
   \langle A B \rangle_{dp}=&  \mathrm{Tr}[(A_{+}^{1}\rho A_{+}^{1}+A_{+}^{2}\rho A_{+}^{2})B+(A_{+}^{2}\rho A_{+}^{1}+A_{+}^{1}\rho A_{+}^{2})B\nonumber\\
   &-(A_{-}^{1}\rho A_{-}^{1}+A_{-}^{2}\rho A_{-}^{2})B-(A_{-}^{1}\rho A_{-}^{2}+A_{-}^{2}\rho A_{-}^{1})B]
\end{align}

However, using the fine-grained degeneracy-breaking projective measurement with four projectors, the joint probability can be written as
\begin{equation}
p_{db}(++)
=
\operatorname{Tr}
\!\left[(A^{1}_{+} \rho A^{1}_{+}+A^{2}_{+} \rho A^{2}_{+})
B_{+}
\right].
\end{equation}
and we derive
\begin{align}\label{von}
   \langle A B \rangle_{db}=&  \mathrm{Tr}[(A_{+}^{1}\rho A_{+}^{1}+A_{+}^{2}\rho A_{+}^{2})B-(A_{-}^{1}\rho A_{-}^{1}+A_{-}^{2}\rho A_{-}^{2})B]\nonumber\\
\end{align}
Comparing Eq. (\ref{Luder}) and Eq. (\ref{von}), we find an interesting relation
\begin{align}
\label{dbdp}
\langle A B \rangle _{db} = &\langle A B\rangle_{dp}- \mathrm{Tr}[(A_{+}^{1}\rho A_{+}^{2}+A_{+}^{2}\rho A_{+}^{1}-A_{-}^{1}\rho A_{-}^{2}-A_{-}^{2}\rho A_{-}^{1})B]
\end{align}
Therefore, the values of $\langle AB \rangle_{db}$ and $\langle A B \rangle_{dp}$ can be different since the second term on the right-hand side of Eq. (\ref{dbdp}) can contribute a non-zero value for a given state and observables.  
\subsection{Violation of TTB }
Now, we consider the LG scenario where the matrices corresponding to the observables $ M_1$, $ M_2$, and $ M_3$ which are mutually commuting. This implies that the measurement is non-disturbing under coarse-grained degeneracy-preserving projective measurement scenario, and no violation of the LG inequality is expected. Surprisingly, by using fine-grained degeneracy-breaking projective measurement, we demonstrate that not only the LG inequality is violated, but TTB is also violated. 

To derive the correlation $\langle M_1 M_2\rangle_{db}$ in fine-grained degeneracy-breaking projective measurement scenario, we consider the following four projectors.
\begin{eqnarray}
    \label{ddb1}
    \nonumber
M_{1+}^1=\frac{1}{4}(\mathbb{I}_{d} + M_{1}+R-S), \ M_{1+}^2=\frac{1}{4}(\mathbb{I}_d +M_1-R+S) \\
M_{1-}^{1}=\frac{1}{4}(\mathbb{I}_{d}-M_1+R+S), \  
M_{1-}^{2}=\frac{1}{4}(\mathbb{I}_{d}-M_1-R-S)
\end{eqnarray}
where  $M_{1+}=M_{1+}^{1}+M_{1+}^{2}$ and $M_{1-}=M_{1-}^{1}+M_{1-}^{2}$ with $M_{1}=M_{1+}-M_{1-}$. Here, the set $\{ M_1, R , S \}$ is mutually commuting with $ R=- M_1 S$. Now, using Eq. (\ref{dbdp}), the correlation can be derived as   
\begin{equation}
\label{hh2}
    \langle M_{1} M_{2} \rangle_{db} = \frac{\langle M_1 M_2 \rangle_{dp}}{2} + \frac{1}{2} \operatorname{Tr} [\rho S(M_1 M_2) S]
\end{equation}
Similarly, 
\begin{equation}
    \langle M_{1} M_{3} \rangle_{db} = \frac{\langle M_1 M_3 \rangle_{dp}}{2} + \frac{1}{2} \operatorname{Tr} [ \rho S(M_1 M_3) S]
\end{equation}

For $\langle M_{2} M_{3} \rangle_{db}$, following Eq. (\ref{ddb1}), we consider four projectors constructed by the commuting set of observables $\{ M_2,  E,  F \}$ with $ F = - M_2 E$. We then have 
\begin{equation}
\langle M_{2} M_{3} \rangle _{db}= \frac{\langle M_2 M_3 \rangle _{dp}}{2} +\frac{1}{2} \operatorname{Tr} [ \rho F(M_2 M_3) F]
    \end{equation}
    
Therefore, from Eq. (\ref{lgi}), we get
\begin{equation}
\label{dbfinal}
K_{db} = \frac{\langle K\rangle_{dp} }{2}+\frac{\operatorname{Tr}[\rho(S(M_1 M_2) S+ F(M_2 M_3) F -S(M_1 M_3) S)]}{2}
  \end{equation}
  Here $\langle K\rangle_{dp}$ is the value of LG functional in coarse-grained degeneracy preserving projective measurement.
  Since the effects corresponding to $M_1$, $M_2$, and $M_3$ are jointly measurable in that scenario, the maximum value of $\langle K\rangle_{dp}$ can be $1$. By suitably choosing the observables $F$ and $S$, if we can find \,$\operatorname{Tr} [\rho(S(M_1 M_2) S)] = \operatorname{Tr} [\rho (F (M_2 M_3) F)] = -\operatorname{Tr} [\rho(S(M_1 M_3) S)]=1$, then this can lead to $K_{db}=2$. 
  
Note that, we have not mentioned the dimension of the system yet. However, a two-qubit realization is as follows. We consider the commuting observables  $ M_{1}=\sigma_{x}\otimes \sigma_{x}$, $ M_{2} = \sigma_{y} \otimes \sigma_{y} $ and $M_{3} = \sigma_{z} \otimes \sigma_{z}$ and the state $|\psi\rangle=\frac{1}{\sqrt{2}}(|01\rangle+|10\rangle)$. By choosing $S=\sigma_{z}\otimes \sigma_{z}$ and $F=\sigma_{y}\otimes  \mathbb{I}_{2}$  we get $K_{db} =2$, violating the TTB bound. Note that, by increasing the number of projectors, the value of $K_{db}$ can be further increased, and for a sufficiently large number of projectors, the value of $K_{db}$ can reach $3$.

 We emphasize that, for the coarse-grained projective measurements of three  mutually commuting observables, the violation of LGI does not occur. But the measurement performed through the fine-grained degeneracy-breaking projectors, injects a form of non-commutativity, leading to the violation of TTB of LG functional. For example, the coarse-grained projectors corresponding to $M_1$, $M_2$ and $M_3$ are commuting. But the fine-grained projectors of $M_2$, in general, do not commute with projectors of $M_1$ and $M_3$.

\subsection{Conceptual shortcoming in TTB violation}

\begin{figure}[h!]
   \centering
 \captionsetup{justification=centering}
 \includegraphics[  width=1.2\columnwidth,  trim=0 0 0 10,  clip]{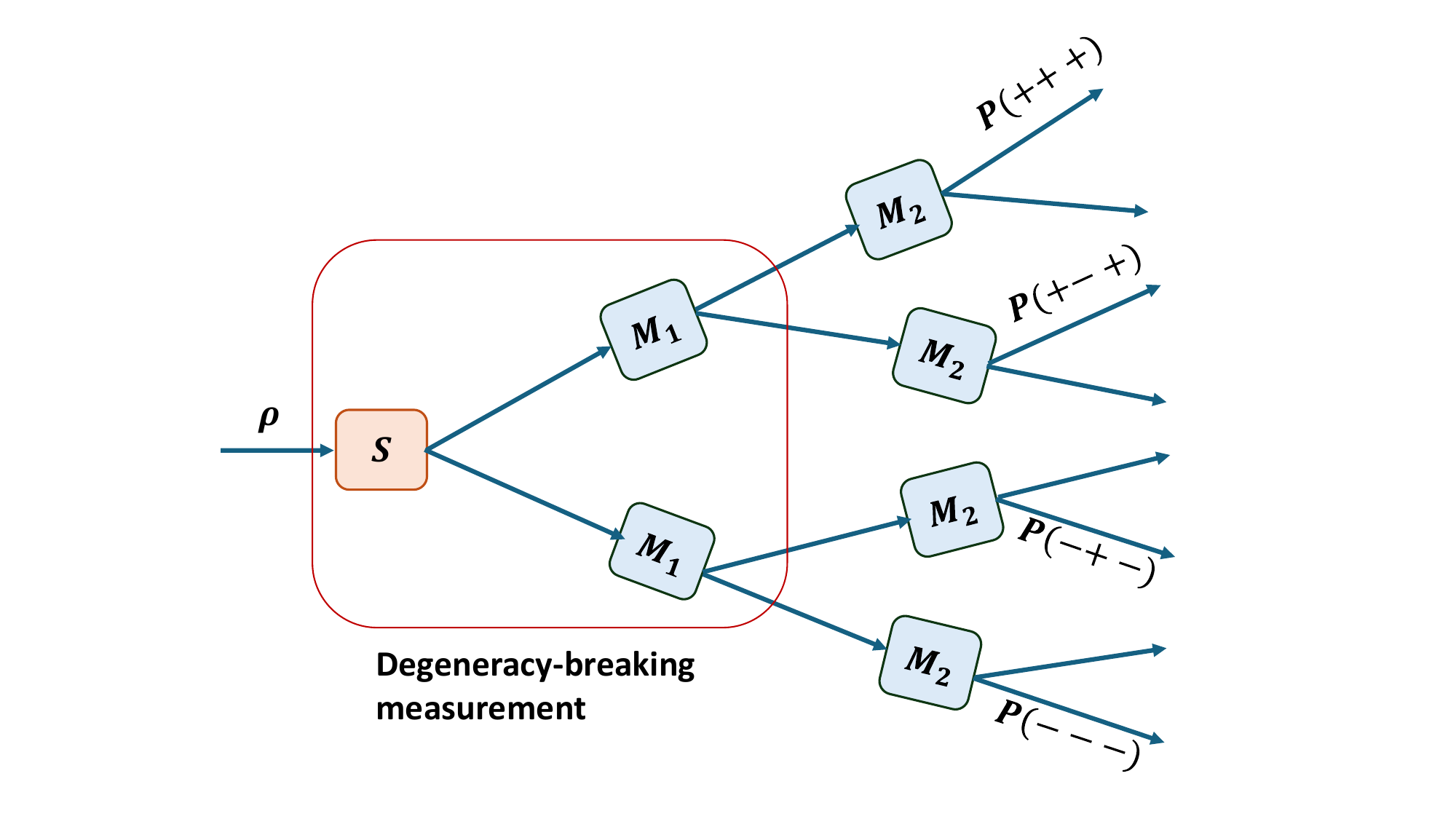}    
\caption{A four-projector degeneracy-breaking measurement scheme. As depicted within the red box, the upper and lower branches after the measurement of $S$ correspond to the post-measurement states $S_+ \rho S _{+}$ and $S_{-} \rho S _{-}$, respectively. The  measurement $M_1$ is performed on each branch which finally implements the degeneracy-breaking measurement.}
\label{fig deg breaking}
\end{figure}

We just demonstrated that, the TTB can be violated if the fine-grained degeneracy-breaking projective measurement is employed. We note that to perform a degeneracy-breaking measurement corresponding to the dichotomic observable $M_1$, one needs a prior measurement \cite{heg,kathakali} of an observable commuting to $M_1$. For example, for the four-projector degeneracy-breaking measurement of $M_1$, a prior commuting measurement of $S$ is performed with $[M_1, S]=0$. From Fig. \ref{fig deg breaking}, we can then write the joint probability for the degeneracy-breaking measurement as   
\begin{align}\label{prob vn}
    p_{db}(++|M_{1},M_{2})=&p(+++|S,M_{1}, M_{2})+p(-++|S,M_{1}, M_{2})\nonumber\\
       =& \mathrm{Tr}[\rho^{\prime} (M_{1+}M_{2+}M_{1+})]
\end{align}
where
\begin{align}
\rho^{\prime}=S_{+}\rho S_{+}+S_{-}\rho S_{-}
\end{align}
with $S_{+}+S_{-}=\mathbb{I}_{d}$. Therefore, the act of degeneracy-breaking measurement leading to calculate the joint probability for a different state $\rho^{\prime}$ instead the initial state $\rho$. The joint probability  for degeneracy-preserving measurement is given by
\begin{align}
      p_{dp}(++|M_{1},M_{2})=\mathrm{Tr}[\rho (M_{1}^{+}M_{2}^{+}M_{1}^{+})]
\end{align}
In general, $p_{db}(++|M_{1},M_{2})\neq p_{dp}(++|M_{1},M_{2})$ except for $\rho = \mathbb{I}_{d}/d$. 

The correlations $\langle M_1 M_2\rangle_{db}$ can then be written as 

\begin{eqnarray}
\label{hh1}
\langle M_1 M_2\rangle_{db} = \frac{\mathrm{Tr}[\rho^{\prime} \{M_1, M_2\}]}{2}
\end{eqnarray}
Note that here $[M_1,M_2]=0$ and hence $\langle M_1 M_2\rangle_{db} = \mathrm{Tr}[\rho^{\prime} M_1 M_2]$. To show similarity with Eq. (\ref{seqq}) we write Eq. (\ref{hh1}) in that form.  By taking $S_{\pm}=(\mathbb{I}_{d} \pm S)/2$, we have $\rho^{\prime}=(\rho+S\rho S)/2$. Putting it in Eq. (\ref{hh1}) we get Eq. (\ref{hh2}) back, as expected.

Now, we raise the crucial point. For deriving the correlations $\langle M_1 M_2\rangle_{db}$ and $\langle M_1 M_3\rangle_{db}$ one performs the measurement on the state $\rho^{\prime}$. However, for the measurement of $\langle M_2 M_3\rangle_{db}$ one needs to consider a different observable $F$ prior to the measurement of $M_{2}$. Then, the correlation $\langle M_2 M_3\rangle_{db}$ is derived for a different state
\begin{align}
\rho^{\prime\prime}=F_{+}\rho F_{+}+F_{-}\rho F_{-}
\end{align}
satisfying $F_{+}+F_{-}=\mathbb{I}_{d}$.

Therefore, by employing degeneracy-breaking measurement, the quantum LG functional becomes

\begin{eqnarray}
    \label{lgi1}
    K_{db} = \langle {M_{1}} {M_{2}}\rangle _{\rho^{\prime}} +\langle {M_{2}} {M_{3}}\rangle _{\rho^{ \prime \prime}}-\langle {M_{1}} {M_{3}}\rangle _{\rho^{\prime}}
\end{eqnarray}
which is different from Eq. (\ref{lgi}) as the three correlations in Eq. (\ref{lgi1}) are derived for two different states $\rho^{\prime}$ and $\rho^{\prime\prime}$. Therefore, it is not surprising to obtain a violation of TTB and even to get the algebraic maximum of the LG functional.

This clearly differs from the conventional LG framework where the correlations in Eq. (\ref{lgi}) were needed to be derived for one state $\rho$. We would like to state that there can be two possible bounds in a macrorealist model depending upon the way one interprets the description of ontic states in a macrorealist model. The case when $\rho'$ and $\rho''$ are macrorealistically equivalent, one gets the macrorealist bound $K\leq 1$, and hence the extreme violation is justified within the LG framework. However, a macrorealist may consider two inequivalent ontic distributions $\mu(\lambda^{\prime})$ and $\mu( \lambda^ {\prime \prime})$ corresponding to two different states $\rho^{\prime}$ and $\rho^{\prime\prime}$. In such a case, for two distinct ontic distributions, the observable $M_2$ can take the value $+1$ in the correlation $\langle M_{1}M_{2}\rangle _{\lambda^{\prime}}$ and $-1$ in $\langle M_2 M_3\rangle _{\lambda''}$, yielding the macrorealist bound of LG functional $K \leq 3$ from Eq. (\ref{lgi1}). Then no violation of LG inequality occurs in a macrorealist model. We leave it for reader which interpretation they choose.


\section{On the violation of TTB that uses superposition of unitaries}
We now focus on a recent work by Chatterjee \emph{et al.} \cite{usha25} that also demonstrated the violation of TTB. They claimed that the superposition of unitaries plays a crucial role in obtaining the extreme violation of the LG inequality. By providing a simple example, we first demonstrate that the unitary superposition is not necessary to obtain their results and then argue why their scenario is beyond the LG framework. 
\subsection{Violation of TTB without superposition of unitaries}
As in \cite{usha25}, we consider a qubit system and choose the initial state \(\rho = \mathbb{I}/2\). Let us now consider the following unitary evolutions \( U(t_1,t_2) = e^{-i g_1 \sigma_{\theta}} \), \( U(t_2,t_3) = e^{-i g_2 \sigma_{\theta}} \), and finally   
\( U(t_1,t_3) = e^{-i (g_1 + g_2) \sigma_{\eta}} \),  
where \( g_1 = \omega (t_2 - t_1) \) and \( g_2 = \omega (t_3 - t_2) \) with $\omega$ being a suitable coupling constant. The operator \(\sigma_{\xi}\) takes the form

\begin{align}
\sigma_{\xi} = \cos{\xi}\,\sigma_x + \sin{\xi}\,\sigma_z
\end{align}
where  $\xi\in\{\theta,\eta\}$. Note here that, 
$U(t_1,t_3)\neq  U(t_2,t_3)  U(t_1, t_2)$ 
unless $\eta=\theta$.

Next, by taking $M_{1} =\sigma_z$\,, we derive the quantum value of the LG expression in Eq. (\ref{lgi}) as
\begin{align}\label{k3}
K_Q =& \cos^{2}g_1 - \cos2\theta\,\sin^{2}g_1
       + \cos^{2}g_2 - \cos2\theta\,\sin^{2}g_2 \nonumber\\
     & - \cos^{2}(g_1+g_2) + \cos2\eta\,\sin^{2}(g_1+g_2)
\end{align}
We consider two cases, when $\eta=\theta$ and $\eta\neq\theta$ and derive the optimal quantum value of $K_{Q}$ in both the cases.

\textbf{Case}- I: When  $\eta = \theta$, i.e., $U(t_1,t_3)=  U(t_2, t_3)  U(t_1, t_2)$. In this case the maximum value of $K_{Q}$ occurs for  $\eta = n\pi$. We find 
\begin{align}
    K_Q = \cos2g_1 + \cos2g_2 - \cos2(g_1+g_2)
\end{align}
The maximum value $(K_Q)_{max}=1.5$ is obtained for $g_1 = g_2 =\pi/6$. This is the TTB of LG functional $K_{Q}$.

\textbf{Case}- II: When $\eta \neq \theta$ i.e., $U(t_1,t_3)\neq  U(t_2, t_3)  U(t_1, t_2)$. Putting again  $\eta = n\pi$, we have
\begin{align}\label{K31}
K_Q = 2 - 2 \cos^{2}\theta \ (\sin^2{g_1}  +  \sin^2{g_2}) -\cos2(g_1+g_2)
\end{align}
On substituting $g_1+g_2=\pi/2$ and $\theta=\pi/2$ in Eq. (\ref{K31}), we get $(K_Q)_{max}=3$, which is the algebraic maximum of $K_{Q}$. \\

\subsection{Conceptual issue in the extreme violation}

We demonstrated above that the superposition of unitaries is not necessary to obtain the extreme violation. In fact, the key ingredient in Chatterjee \emph{et al.} \cite{usha25} is the requirement  $U(t_1,t_3)\neq U(t_2, t_3)U(t_1, t_2)$. Note that the correlation $\langle M_1 M_2\rangle$ in Eq. (\ref{lgi}) is derived by considering  $M_{2}=U ^{\dagger} (t_1,t_2) M_{1} U (t_1,t_2)$. Now, for the correlation $\langle M_2 M_3\rangle$ we take $M_{3}=U ^{\dagger} (t_2,t_3) M_{2}U(t_2,t_3)$ which is equal to 
\begin{eqnarray}
M_{3}=U ^{\dagger} (t_1,t_2)U ^{\dagger} (t_2,t_3) M_{1} U(t_2,t_3) U(t_1,t_2)
\end{eqnarray} 
However, to derive the correlations $\langle M_1 M_3\rangle$, Chatterjee \emph{et al.} \cite{usha25} assumed  $U(t_1,t_3)\neq U(t_2, t_3) U(t_1,t_2)$. This simply means $U ^{\dagger} (t_1,t_3)M_{1}U (t_1,t_3)\equiv M_{3}^{\prime}$. Therefore, $M_{3}$ and $M_{3}^{\prime}$ are different observables.  In Chatterjee \emph{et al.} \cite{usha25}, the LG functional in Eq. (\ref{lgi}) is then revised to
\begin{eqnarray}
\label{lgi111}
K^{\prime}_{Q}=\langle {M_{1}} {M_{2}}\rangle_{\rho} +\langle {M_{2}} {M_{3}}\rangle_{\rho}-\langle {M_{1}} {M_{3}^{\prime}}\rangle_{\rho}
\end{eqnarray}
which is also different from Eq. (\ref{lgi}) as $M_{3}\neq M_{3}^{\prime}$.  Therefore, the scenario considered in Chatterjee \emph{et al.} \cite{usha25} is outside the conventional LG framework as described in Fig. \ref{fig:LG}. It is straightforward to check that in a macrorealist model $K^{\prime} = 3$ can be obtained and hence reproduces $(K_Q)_{max}$.

 \section{Summary and discussion}
 
In sum, we critically examined the violation of TTB of LG expression in Eq. (\ref{lgi}) claimed by Budroni and Emary \cite{budroni14} and by  Chatterjee \emph{et al.} \cite{usha25}. In  \cite{budroni14}, the violation of TTB was demonstrated by invoking degeneracy-breaking measurement and in \cite{usha25} by considering the evolution between two measurements to be the superposition of unitaries. We first provided simpler examples to demonstrate the results of \cite{budroni14, usha25}. Instead of non commuting observables taken in \cite{budroni14}, we reproduce the TTB violation by considering three observables which are mutually commuting in the coarse-grained L\"uders measurement scenario. We also showed that the results of \cite{usha25} can be obtained without the superposition of unitaries.
 
Further we argue that the frameworks invoked to exhibit TTB violations differ from the conventional LG framework. Specifically, we demonstrated that the action of degeneracy-breaking measurement employed in \cite{budroni14} leads to the LG functional in Eq. (\ref{lgi1}) where three correlations are effectively derived for two different initial states.  Furthermore, we showed that the unitary evolutions taken in \cite{usha25} lead to an LG functional in Eq. (\ref{lgi111}) where three correlations are effectively computed for four observables. Therefore, in the framework proposed in\cite{budroni14,usha25}, the appropriate macrorealist bound becomes $K\leq 3$, implying that there is no violation of the LG inequality. Hence, the TTB violations reported in \cite{budroni14,usha25} have only a limited bearing on the conventional LG framework and, consequently, do not constitute a test of the incompatibility between macrorealism and quantum theory.

\emph{Acknowledgments:-}
A.B. acknowledges funding from the DST-Inspire fellowship No. DST/INSPIRE
Fellowship/2023/IF210650, Govt. of India. A.K.P. acknowledges the support from the research grant I-HUB/PDF/2022-23/06, Government of India.


\begin{thebibliography}{99}




\bibitem{sch} E. Schrödinger, (1935). The present situation in quantum mechanics. \href{https://doi.org/10.1007/BF01491891}{Naturwissenschaften \textbf{23}, 807–812 (1935)}.


\bibitem{zeh} H. D. Zeh, On the interpretation of measurement in quantum theory, \href{https://doi.org/10.1007/BF00708656}{ Found. Phys. \textbf{1}, 69–76 (1970)}.


\bibitem{zur} W. H. Zurek, Environment-induced superselection rules, \href{ https://doi.org/10.1103/PhysRevD.26.1862}{Phys. Rev. D \textbf{26}, 1862 (1982)}.



\bibitem{bruk} J. Kofler and C. Brukner, Classical World Arising out of Quantum Physics under the Restriction of Coarse-Grained Measurements, \href{ https://doi.org/10.1103/PhysRevLett.99.180403}{Phys. Rev. Lett. \textbf{99}, 180403 (2007)}.


\bibitem{ghc86}G. C. Ghirardi, A. Rimini and T. Weber, Unified dynamics for microscopic and macroscopic systems, \href{ https://doi.org/10.1103/PhysRevD.34.470}{Phys. Rev. D \textbf{34}, 470 (1986)}.

\bibitem{ghc03} A. Bassi and G. Ghirardi, Dynamical reduction models, \href{ https://doi.org/10.1016/S0370-1573(03)00103-0}{Phys. Rep. \textbf{379}, 257–426 (2003)}.


\bibitem{arndt} M. Arndt, O. Nairz, J. v. Andreae, C. Keller, G. Zouw, A. Zeilinger, Wave–particle duality of C60 molecules, \href{https://www.nature.com/articles/44348}{Nature, \textbf{401}, 680-682 (1999)}. 



\bibitem{gerlich} S. Gerlich, S. Eibenberger, M. Tomandl, S. Nimmrichter, K. Hornberger, P. J. Fagan, J. Tüxen, M. Mayor and M. Arndt, Quantum interference of large organic molecules, \href{ https://doi.org/10.1038/ncomms1263}{Nat. Commun. \textbf{2}, 263 (2011)}.




\bibitem{Schrinski} B. Schrinski, Y. Yang, U. v. Lüpke, M. Bild, Y. Chu, K. Hornberger, S. Nimmrichter and M. Fadel, Macroscopic Quantum Test with Bulk Acoustic Wave Resonators, \href{ https://doi.org/10.1103/PhysRevLett.130.133604 }{Phys. Rev. Lett. \textbf{130}, 133604 (2023)}.


\bibitem{ped} S. Pedalino, B. E. Ramírez-Galindo, R. Ferstl, K. Hornberger, M. Arndt and S. Gerlich, Probing quantum mechanics with nanoparticle matter-wave interferometry, \href{ https://doi.org/10.1038/s41586-025-09917-9}{ Nature \textbf{649}, 866–870 (2026)}. 



\bibitem{triel} B. S. Cirel'son, Quantum generalizations of Bell's inequality. \href{https://doi.org/10.1007/BF00417500}{Lett. Math. Phys. \textbf{4}, 93–100 (1980)}. 
\bibitem{bell64}J. S. Bell, On the Einstein Podolsky Rosen paradox, \href{ https://doi.org/10.1103/PhysicsPhysiqueFizika.1.195 }{Phys. Phys. Fiz. \textbf{1}, 195, (1964)}.

\bibitem{leggett85}A. J. Leggett and A. Garg, Quantum mechanics versus macroscopic realism: Is the flux there when nobody looks?, \href{https://doi.org/10.1103/PhysRevLett.54.857}{Phys. Rev. Lett. \textbf{54}, 857 
(1985)}.

\bibitem{leggett02}A. J. Leggett, Testing the limits of quantum mechanics: motivation, state of play, prospects, \href{https://iopscience.iop.org/article/10.1088/0953-8984/14/15/201}{ J. Phys. Condens. Matter \textbf{14}, R415 (2002)}.

\bibitem{leggett08} A. J. Leggett, Realism and the physical world, \href{https://iopscience.iop.org/article/10.1088/0034-4885/71/2/022001}{Rep. Prog. Phys. \textbf{71} 022001 (2008)}.




\bibitem{Emary13}C. Emary, N. Lambert and F. Nori, Leggett–Garg inequalities, \href{https://iopscience.iop.org/article/10.1088/0034-4885/77/1/016001}{ Rep. Prog. Phys. \textbf{77} 016001, (2014)}.

\bibitem{halli16} J. J. Halliwell, Leggett-Garg inequalities and no-signaling in time: A quasiprobability approach, \href{https://doi.org/10.1103/PhysRevA.93.022123}{ Phys. Rev. A, \textbf{93}, 022123 (2016)}.

\bibitem{halli17} J. J. Halliwell, Comparing conditions for macrorealism: Leggett-Garg inequalities versus no-signaling in time, \href{ https://doi.org/10.1103/PhysRevA.96.012121}{ Phys. Rev. A \textbf{96}, 012121 (2017)}.

\bibitem{halli19} J. J. Halliwell, Leggett-Garg tests of macrorealism: Checks for noninvasiveness and generalizations to higher-order correlators, \href{https://doi.org/10.1103/PhysRevA.99.022119}{Phys. Rev. A \textbf{99}, 022119 (2019)}. 

\bibitem{halli19a}J. J. Halliwell and C. Mawby, Fine's theorem for Leggett-Garg tests with an arbitrary number of measurement times, \href{https://doi.org/10.1103/PhysRevA.100.042103}{Phys. Rev. A \textbf{100}, 042103 (2019)}.


\bibitem{halli22} C. Mawby and J. J. Halliwell, Leggett-Garg tests for macrorealism in the quantum harmonic oscillator and more general bound systems, \href{ https://doi.org/10.1103/PhysRevA.105.022221}{Phys. Rev. A \textbf{105}, 022221 (2022)}.

\bibitem{halli20} J. J. Halliwell and C. Mawby, Conditions for macrorealism for systems described by many-valued variables, \href{ https://doi.org/10.1103/PhysRevA.102.012209}{Phys. Rev. A \textbf{102}, 012209, 2020}.

\bibitem{brukner14} B. Dakić, T. Paterek and Č. Brukner, Density cubes and higher-order interference theories, \href{ https://iopscience.iop.org/article/10.1088/1367-2630/16/2/023028}{New J. Phys. \textbf{16} 023028, (2014)}.

\bibitem{wild} M. Wild and A. Mizel, Addressing the Clumsiness Loophole in a Leggett-Garg Test of Macrorealism, \href{https://link.springer.com/article/10.1007/s10701-011-9598-4}{ Found. Phys. \textbf{42}, 256-265 (2012)}.

\bibitem{maroney} O. J. E. Maroney and C. G. Timpson, Quantum-vs. Macro- Realism: What does the Leggett-Garg Inequality actually test? \href{ https://doi.org/10.48550/arXiv.1412.6139}{arXiv:1412.6139 (2014)}.


\bibitem{clemente15}L. Clemente and J. Kofler, Necessary and sufficient conditions for macroscopic realism from quantum mechanics, \href{https://doi.org/10.1103/PhysRevA.91.062103}{Phys. Rev. A, \textbf{91}, 062103 (2015)}.

\bibitem{clemente16}L. Clemente and J. Kofler, No Fine Theorem for Macrorealism: Limitations of the Leggett-Garg Inequality, \href{https://doi.org/10.1103/PhysRevLett.116.150401}{ Phys. Rev. Lett. \textbf{116}, 150401 (2016)}.

\bibitem{pan17} S. Kumari and A. K. Pan, Probing various formulations of macrorealism for unsharp quantum measurements, \href{https://doi.org/10.1103/PhysRevA.96.042107}{ Phys. Rev. A \textbf{96}, 042107 (2017)}. 

\bibitem{akp23} A. K. Pan, Leggett-Garg test of macrorealism using indefinite causal order of measurements, \href{ https://doi.org/10.1016/j.physleta.2023.128898 }{ Phys. Lett. A \textbf{478}, 128898 (2023)}.

\bibitem{weak value and MR akp20} A. K. Pan, Interference experiment, anomalous weak value, and Leggett-Garg test of macrorealism, \href{ https://doi.org/10.1103/PhysRevA.102.032206}{Phys. Rev. A \textbf{102}, 032206 (2020)}.

\bibitem{budroni23} G. Vitagliano and C. Budroni, Leggett-Garg macrorealism and temporal correlations, \href{https://doi.org/10.1103/PhysRevA.107.040101}{Phys. Rev. A \textbf{107}, 040101 (2023)}.

\bibitem{budroni13} C. Budroni, T. Moroder, M. Kleinmann and O. Gühne, Bounding Temporal Quantum Correlations, \href{https://doi.org/10.1103/PhysRevLett.111.020403}{Phys. Rev. Lett. \textbf{111}, 020403 (2013)}.

\bibitem{budroni14} C. Budroni and C. Emary, Temporal Quantum Correlations and Leggett-Garg Inequalities in Multilevel Systems, \href{https://doi.org/10.1103/PhysRevLett.113.050401}{Phys. Rev. Lett. \textbf{113}, 050401 (2014)}.





\bibitem{swati17} S. Kumari and A. K. Pan, Inequivalent Leggett-Garg inequalities, \href{https://iopscience.iop.org/article/10.1209/0295-5075/118/50002}{ EPL \textbf{118}, 50002 (2017)}.

\bibitem{kumari18} A. Kumari, Md. Qutubuddin and A. K. Pan, Violation of the Lüders bound of macrorealist and noncontextual inequalities, \href{https://doi.org/10.1103/PhysRevA.98.042135}{Phys. Rev. A \textbf{98}, 042135 (2018)}.


\bibitem{pan18} A. K. Pan, Md. Qutubuddin and S. Kumari, Quantum violation of variants of Leggett-Garg inequalities up to the algebraic maximum for a qubit system, \href{https://doi.org/10.1103/PhysRevA.98.062115}{Phys. Rev. A, \textbf{98}, 062115 (2018)}.






\bibitem{knee} G. C. Knee, S. Simmons, E. M. Gauger, J. J. L. Morton, H. Riemann, N. V. Abrosimov, P. Becker, H. -J.  Pohl, K. M. Itoh, M. L.W. Thewalt, G. A. D. Briggs and S. C. Benjamin, Violation of a Leggett–Garg inequality with ideal non-invasive measurements, \href{ https://www.nature.com/articles/ncomms1614}{ Nat. Commun. \textbf{3}, 606 (2012)}.

\bibitem{robens} C. Robens, W. Alt, D. Meschede, C. Emary and A. Alberti, Ideal Negative Measurements in Quantum Walks Disprove Theories Based on Classical Trajectories, \href{https://doi.org/10.1103/PhysRevX.5.011003}{ Phys. Rev. X \textbf{5}, 011003 (2015)}.


\bibitem{Goggin} M. E. Goggin, M. P. Almeida, M. Barbieri and G. J. Pryde, Violation of the Leggett–Garg inequality with weak measurements of photons, \href{https://doi.org/10.1073/pnas.1005774108}{Proc. Natl. Acad. Sci. U.S.A. \textbf{108}, 1256-1261 (2011)}.

\bibitem{pala} A. Palacios-Laloy, F. Mallet, F. Nguyen, P. Bertet, D. Vion, D. Esteve and A. N. Korotkov, Experimental violation of a Bell’s inequality in time with weak measurement, \href{https://www.nature.com/articles/nphys1641}{Nat. Phys. \textbf{6}, 442-447 (2010).}
\bibitem{tusun20} M. Tusun, W. Cheng, Z. Chai, Y. Wu, Y. Wang, X. Rong, and J. Du, Experimental violations of Leggett-Garg inequalities up to the algebraic maximum for a photonic qubit, \href{ https://doi.org/10.1103/PhysRevA.102.022214}{Phys. Rev. A \textbf{102}, 022214 (2020)}.


\bibitem{tusun22} M. Tusun, W. Cheng, Z. Chai, Y. Wu, Y. Wang, X. Rong and J. Du, Experimental violation of the Leggett-Garg inequality with a single-spin system, \href{ https://doi.org/10.1103/PhysRevA.105.042613}{Phys. Rev. A \textbf{105}, 042613 (2022)}.


\bibitem{laflamme} H. Katiyar, A. Brodutch, D. Lu and R. Laflamme, Experimental violation of the Leggett–Garg inequality in a three-level system, \href{https://doi.org/10.1088/1367-2630/aa5c51}{New J. Phys. \textbf{19}, 023033 (2017)}.

\bibitem{wang17} K. Wang, C. Emary, X. Zhan, Z. Bian, J. Li and P. Xue, Enhanced violations of Leggett-Garg inequalities in an experimental three-level system, \href{ https://doi.org/10.1364/OE.25.031462}{ Opt. Express \textbf{25}, 31462–31470 (2017)}.

\bibitem{zhan23} T. Zhan, C. Wu, M. Zhang, Q. Qin, X. Yang, H. Hu, W. Su, Jie Zhang, T. Chen, Y. Xie, W. Wu and P. Chen, Experimental violation of the Leggett-Garg inequality in a three-level trapped-ion system, \href{https://doi.org/10.1103/PhysRevA.107.012424}{Phys. Rev. A \textbf{107}, 012424 (2023)}.

\bibitem{sponar} E. Kreuzgruber, R. Wagner, N. Geerits, H. Lemmel and S. Sponar, Violation of a Leggett-Garg Inequality Using Ideal Negative Measurements in Neutron Interferometry, \href{ https://doi.org/10.1103/PhysRevLett.132.260201}{Phys. Rev. Lett. \textbf{132}, 260201 (2024)}.

\bibitem{usha25} A. Chatterjee, H. S. Karthik, T. S. Mahesh and A. R. Usha Devi, Extreme Violations of Leggett-Garg Inequalities for a System Evolving under Superposition of Unitaries, \href{https://doi.org/10.1103/vydp-9qqq}{Phys. Rev. Lett. \textbf{135}, 220202 (2025)}.


\bibitem{Zhang} C.-W. Wu, M.-C. Zhang, Y.-L. Zhou, T. Chen, R. Huang, Y. Xie, W.-b. Su, B.-Q. Ou, W. Wu, A. Miranowicz, F. Nori, J. Zhang, H. Jing, and P.-X. Chen, Observation of quantum temporal correlations well beyond Lüders bound, \href{  https://doi.org/10.1103/PhysRevResearch.7.013058}{Phys. Rev. Research \textbf{7}, 013058 (2025)}.



\bibitem{kathakali} A. K. Pan and K. Mandal, Quantum Contextuality for a Three-Level System Sans Realist Model, \href{ https://doi.org/10.1007/s10773-016-2974-2}{Int J Theor Phys \textbf{55}, 3472–3478 (2016)}.

\bibitem{heg} G. C. Hegerfeldt and R. S. Mayato, Discriminating between the von Neumann and Lüders reduction rule, \href{ https://doi.org/10.1103/PhysRevA.85.032116}{ Phys. Rev. A \textbf{85}, 032116 (2012)}.







\end{thebibliography}
\end{document}